\documentclass[conference]{IEEEtran}
\IEEEoverridecommandlockouts
\usepackage{cite}
\usepackage{amsmath,amssymb,amsfonts}
\usepackage{algorithmic}
\usepackage[table,xcdraw]{xcolor}
\usepackage{graphicx}
\usepackage{lipsum}
\usepackage{textcomp}
\ifCLASSOPTIONcompsoc
    \usepackage[caption=false, font=normalsize, labelfont=sf, textfont=sf]{subfig}
\else
\usepackage[caption=false, font=footnotesize]{subfig}
\fi
\usepackage{xcolor}
\def\BibTeX{{\rm B\kern-.05em{\sc i\kern-.025em b}\kern-.08em
    T\kern-.1667em\lower.7ex\hbox{E}\kern-.125emX}}
\begin{document}

\title{Analog Weights in ReRAM DNN Accelerators\\
}

\author{Jason~K.~Eshraghian$^*$, Sung-Mo Kang$^\dagger$, Seungbum Baek$^\ddagger$, Garrick Orchard$^\mathsection$, Herbert Ho-Ching Iu$^*$, Wen Lei$^*$\\
\IEEEauthorblockA{$^*$\textit{School of Electrical, Electronic and Computer Engineering, University of Western Australia, Crawley, WA 6009 Australia}}
\IEEEauthorblockA{$^\dagger$\textit{Baskin School of Engineering, University of California, Santa Cruz, Santa Cruz, CA 95064 USA}}
\IEEEauthorblockA{$^\ddagger$\textit{College of Electrical and Computer Engineering, Chungbuk National University, Cheongju 362763, South Korea}}
\IEEEauthorblockA{$^\mathsection$\textit{Temasek Laboratories and Singapore Institute for Neurotechnology, National University of Singapore, Singapore 117411}}}

        
\maketitle

\begin{abstract}
Artificial neural networks have become ubiquitous in modern life, which has triggered the emergence of a new class of application specific integrated circuits for their acceleration. ReRAM-based accelerators have gained significant traction due to their ability to leverage in-memory computations. In a crossbar structure, they can perform multiply-and-accumulate operations more efficiently than standard CMOS logic. By virtue of being resistive switches, ReRAM switches can only reliably store one of two states. This is a severe limitation on the range of values in a computational kernel. This paper presents a novel scheme in alleviating the single-bit-per-device restriction by exploiting frequency dependence of \textit{v-i} plane hysteresis, and assigning kernel information not only to the device conductance but also partially distributing it to the frequency of a time-varying input.\\
We show this approach reduces average power consumption for a single crossbar convolution by up to a factor of $\times$16 for an unsigned 8-bit input image, where each convolutional process consumes a worst-case of 1.1mW, and reduces area by a factor of $\times$8, without reducing accuracy to the level of binarized neural networks. This presents a massive saving in computing cost when there are many simultaneous in-situ multiply-and-accumulate processes occurring across different crossbars.
\end{abstract}

\section{Introduction}
The recent rise of IoT and the proliferation of cheap sensors has brought with it an explosion of data requiring inference. Inference hardware is currently a bottleneck in the data processing pipeline, spawning the development of application-specific integrated circuits (ASICs) also known as inference accelerators. The dominant computation during inference in deep neural networks (DNNs) and convolutional neural networks (CNNs) is dot-product multiplication, which translates simply to the multiply-and-accumulate (MAC) operation. This can be expensive to implement and requires numerous logic gates.

Memristive crossbar arrays simplify the hardware mapping of DNN algorithms in two ways: 1) they require less devices per MAC operation; only one single memristor is needed per multiply operation as opposed to potentially over 40 transistors in CMOS for a binary multiplier, and 2) decreasing the physical distance between memory (kernel storage) and computation, thus reducing delay and vulnerability to signal degradation. The use of memristor-based AI accelerators show promise as they enable computation-in-memory architecture for data-intensive applications.

Computational kernels in DNNs require more than two bits per element for high accuracy, but multi-bit memristive crossbars remain highly experimental. This brief presents a novel frequency-dependent scheme to produce analog-valued weights from a single-bit memristor, thus increasing accuracy, and removing the need to distribute computation across various bit-lines which reduces current sneak path and line resistance effects. The use of a DC driving voltage (for inference), as often implicated, is unfeasible as the \textit{v-i} curve converges to a single-valued function and behaves as a non-linear resistor \cite{b1}. This eliminates any switching effect, and necessitates a time-varying input to maintain bi-valued hysteresis which is poorly addressed in the literature. Furthermore, kernels and weights conventionally take on any number of values -- not just `on' and `off'.

By exploiting frequency dependency of memristor conductance, we show it is possible to map a continuous range of weights that are assignable to memristors. The weights are expressed as not only a function of an input voltage pulse (as has been demonstrated in the past) but also as a function of input frequency. Section II describes crossbar operation in DNNs and CNNs to generate the MAC operation, and by extension, 2D convolution. Section III presents a look-up table that maps device conductances to driving frequency, and Section IV provides the derivation of  limitations on the driving input. These analytical results are then used in Section V for experimental verification, where we apply our proposed method to a noisy 128$\times$128 input image through a Gaussian convolution filter for image smoothing. Provided that our method is capable of convolutional image processing, then it will also be able to perform inference in DNNs. Discussions that follow show that the presented frequency-dependent method can reduce crossbar area down to 12.5\% of the original size for an unsigned 8-bit image, with associated power savings that are also characterized and compared against other state-of-the-art methods of image filtering in crossbars. This is all performed without the need to reduce the system down to a binarized NN, which is associated with sacrificing accuracy in more complex processes \cite{b2}. Furthermore, this scheme also enables bit-lines of crossbars to be allocated to other tasks such as the processing of different channels, which thus speeds up the pipeline.
\section{Crossbar Operation}
As shown in Fig.~1(a), a single neuron in an artificial neural network accepts $m+1$ inputs with signals $x_0$ through $x_m$ and weights $w_0$ through $w_m$. The output of a neuron is represented by:
\begin{equation}
y=\sigma \Big(\sum_{j=0}^{m}w_jx_j\Big),	\label{eq1}
\end{equation}
where $\sigma$ is some activation function, in this case the sigmoid function. The term in parentheses is a MAC operation, which can be computed using a resistive switching crossbar \cite{b3,b4,b5,b6,b7,b8,b9,b10,b11a} by mapping weight $w_j$ to conductances $G_j$, and inputs $x_j$ to voltages $V_j$.
The total current at the output of each bit-line is a summation of the current that each memristor within a single column draws from its corresponding input. The summation of current from Fig.~1(b) that is passed through an activation unit is congruous to $y$ from (1). The mapping from the artificial neuron to crossbar architecture can be discerned on inspection of Fig.~1, in mapping neural inspired image processing into hardware \cite{b11,b12}.

Weights $w_j$ are pre-trained to take on any range of values, whereas $G_j$ is typically restricted to either a high or low value. In the following section, we formalize a way to circumvent this limitation without the use of multi-level memristors by using a half-sine pulse as input, and varying the width (i.e., frequency content) of the pulse.
\begin{figure}[t]
\centerline{\includegraphics[scale=0.84]{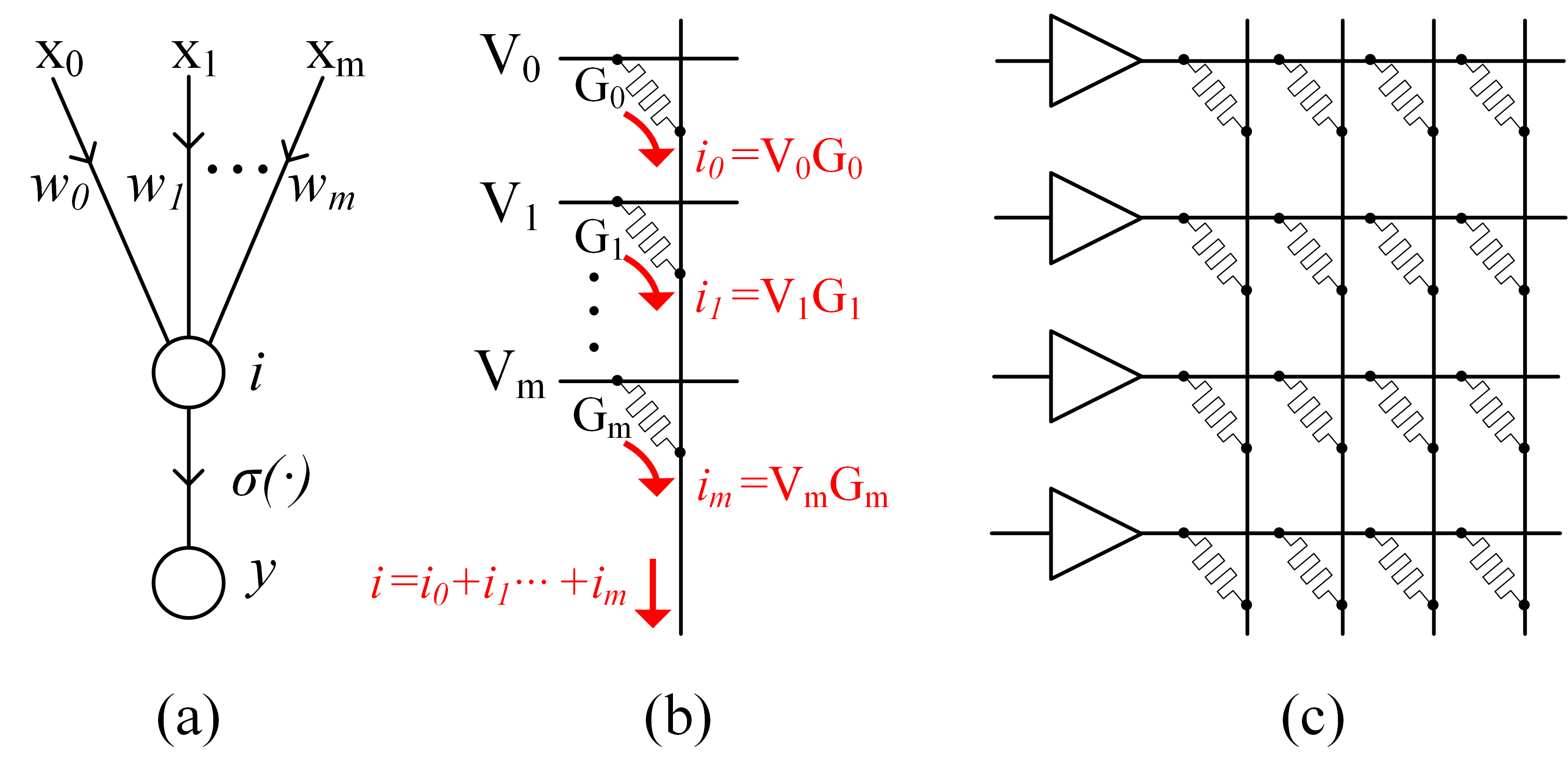}}
\caption{Hardware mapping of neural networks. (a) Artificial neuron model. (b) Multiply-Accumulate operation. (c) Vector-Matrix multiplier.}
\label{fig}
\end{figure}
\section{Mapping Weight to Conductance}
The relationship between frequency and conductance is device-dependent. In this work we use a \textit{Ag}-chalcogenide memristor made up of \textit{GeSeSn-W} \cite{b13} to generate an experimentally verified look-up table. This table is used to select appropriate frequencies for a given conductance (or equivalently, kernel weight). 
The \textit{v-i} characteristic curves are shown in Fig.~2(a), conductance is plotted as a function of frequency in Fig.~2(b), and tabulated in Table I.\footnote{The plots display averaged values across 5 trials at each given frequency. Device reliability will affect the range of weights that can be represented.}

\begin{figure}[t]
\centerline{\includegraphics[scale=0.85]{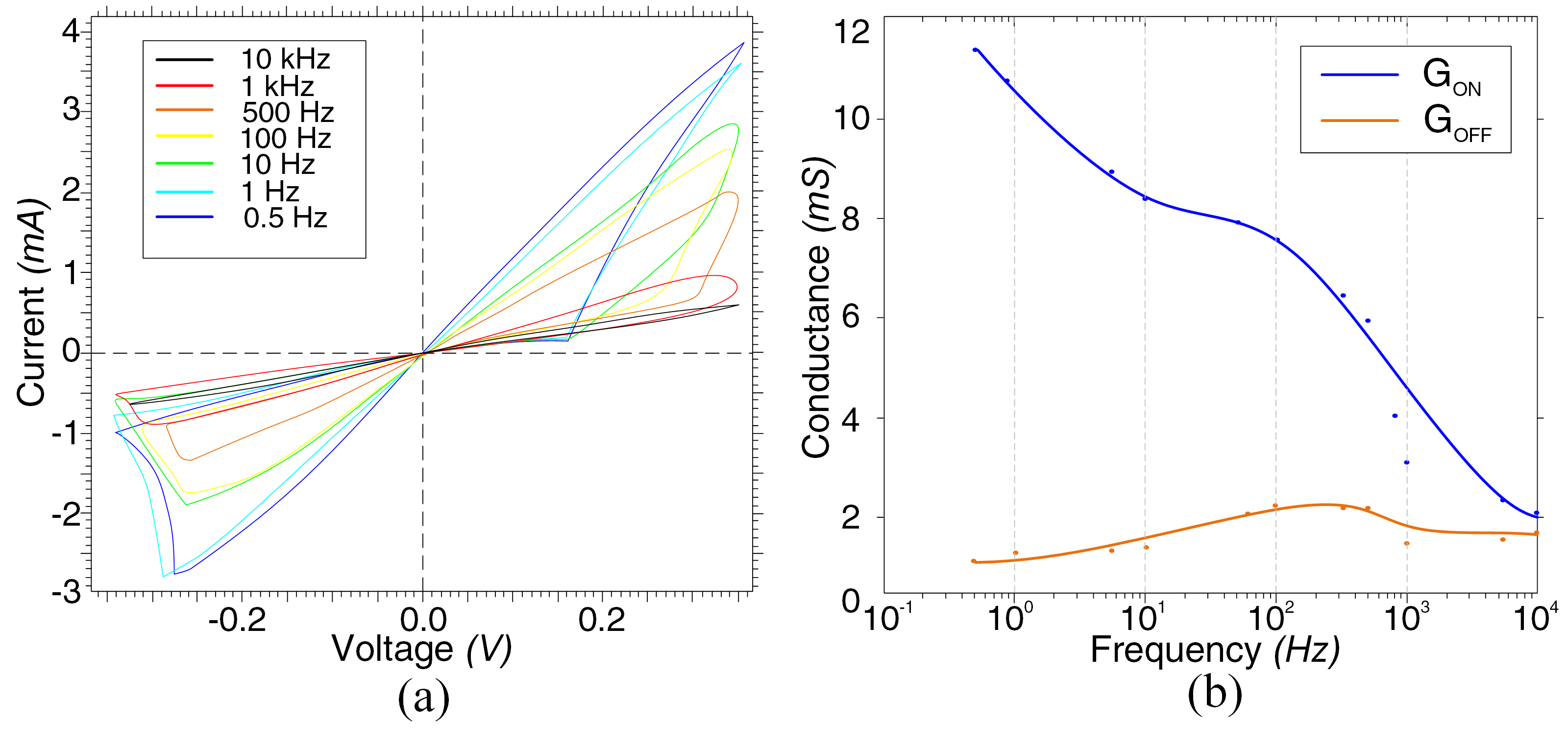}}
\caption{Experimental characteristics of a \textit{GeSeSn-W} memristor (a) \textit{v-i} plane across varying frequencies. (b) Frequency-dependent conductances.}
\label{fig2}
\end{figure}
\begin{table}[t]
\centering\caption{Conductance Look-up Table for \textit{GeSeSn-W} Memristor}
\begin{tabular}{|c|c|c|}
\hline
\rowcolor[HTML]{EFEFEF} 
{Freq. [Hz]} & {$G_{OFF}$ [mS]} & {$G_{ON}$ [mS]} \\ \hline
\rowcolor[HTML]{FFFFFF} 
10k                 & 1.71                                    & 2.10                                   \\ \hline
\rowcolor[HTML]{FFFFFF} 
1k                  & 1.49                                    & 3.13                                   \\ \hline
\rowcolor[HTML]{FFFFFF} 
750                 & 1.56                                    & 4.20                                  \\ \hline
\rowcolor[HTML]{FFFFFF} 
500                 & 2.20                                    & 5.97                                   \\ \hline
\rowcolor[HTML]{FFFFFF} 
100                 & 2.26                                    & 7.60                                   \\ \hline
\rowcolor[HTML]{FFFFFF} 
10                  & 1.4                                     & 8.40                                    \\ \hline
\rowcolor[HTML]{FFFFFF} 
1                   & 1.32                                    & 10.8                                   \\ \hline
\rowcolor[HTML]{FFFFFF} 
0.5                 & 1.15                                    & 11.4                                   \\ \hline
\end{tabular}
\end{table}
Furthermore, Fig. 2(b) depicts the convergence of $G_{ON}$ and $G_{OFF}$ as driving frequency increases, i.e., above a certain frequency the \textit{v-i} characteristic curve degenerates to a single-valued monotonically increasing function. Note that some conductances are repeated across varying frequencies. For example, a frequency of 480 Hz results in $G_{OFF}$ of approximately 2.10mS, and a frequency of 10k Hz results in an approximately identical value of $G_{ON}$. In such duplicative cases there are multiple choices of frequency: a higher frequency will generate faster output, and a lower frequency will dissipate less power. Whether to prioritize speed or power is determined by user requirements. With the frequency-dependent look-up table generated, we must now consider the necessary conditions on amplitude and phase conditions of the input to successfully perform convolutional operations.

\section{Experimental Setup}
The input voltage must be time-varying to ensure a bi-valued conductance function. Here, we use half-sine pulses of duration $T/2s$. Graphically, there is no switching between \textit{v-i} hysteresis branches of resistance during inference -- only partial switching by altering the gradient of the low resistance branch. 

The use of varying frequencies across inputs results in timing mismatch. This is qualitatively depicted in Fig.~3 for demonstration purposes, where $T_j$ is the period of the $j^{th}$ input, $T_{MAX}$ is the period of the widest pulse, $\phi_j$ is the phase-shift of the $j^{th}$ input and $I_{OUT}(t)$ can be calculated using (1) without the activation, and $I_{peak}$ corresponds to the required sum of current amplitudes for MAC computation.

By reference to Fig.~3, without a phase-shift of $\phi_j$, the peaks of $V_j$ and $V_{MAX}$ do not align, which results in current output whose peak does not reflect the MAC operation. The maximum values of each row input must occur simultaneously in order for the peak value of output current to perform MAC. This requires a phase-shift to be introduced in order to align the peak of all row inputs with the maximum period used. In this case, $T_{MAX} = 2s$ as per Table I, and $V_{MAX}$ occurs at $t=T_{MAX}/4s$. The phase for the $j^{th}$ input is therefore a time-varying voltage $V_j$ for a given period $T_j$:
\begin{equation}
\phi_j=\frac{\pi}{2}\big(1-\frac{T_{MAX}}{T_j}\big)\label{eq2}
\end{equation}
\begin{figure}[t]
\centerline{\includegraphics[scale=0.6]{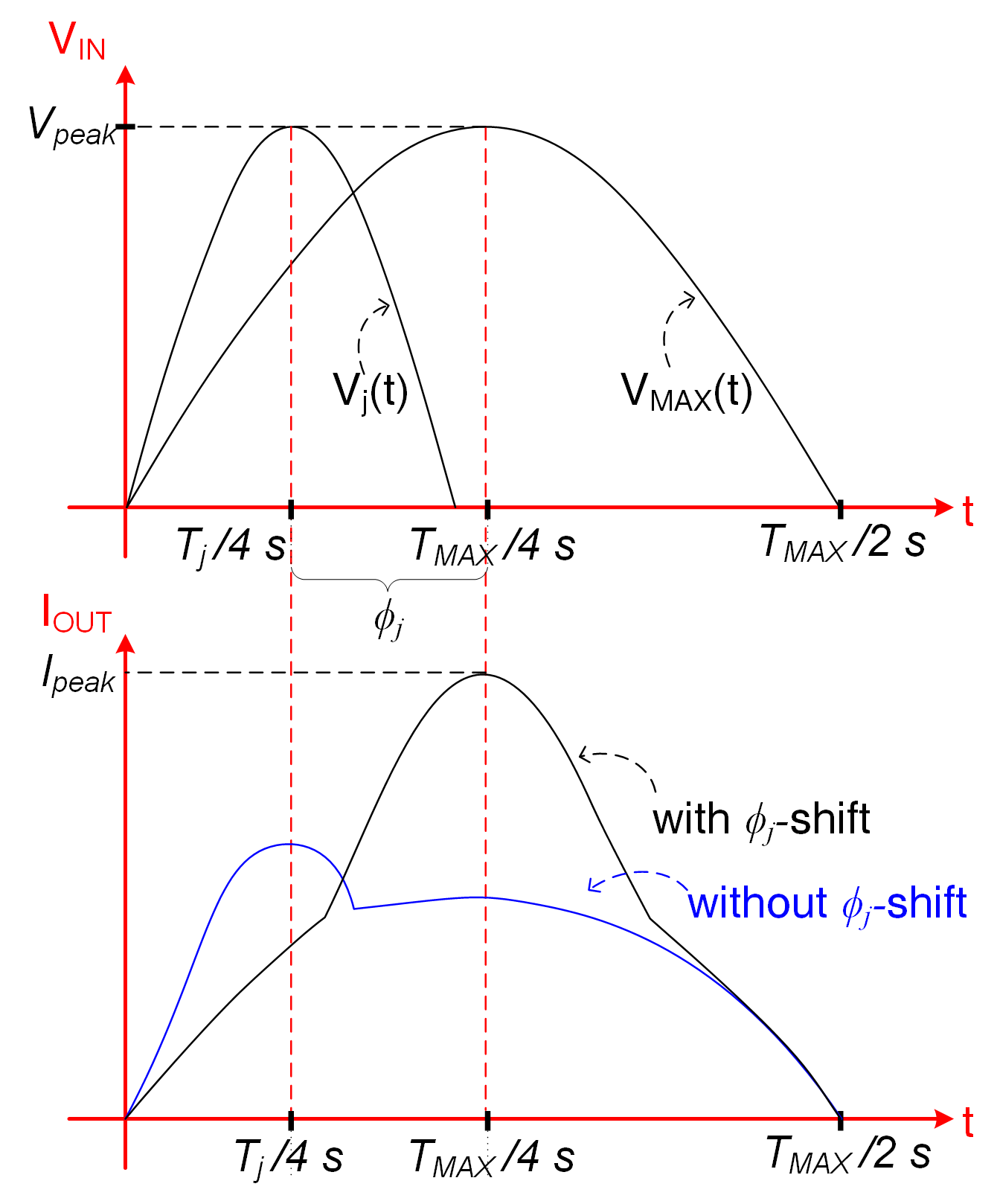}}
\caption{A qualitative depiction of phase-shift calculation for input alignment.}
\label{fig3}
\end{figure}
\noindent This results in a current waveform corresponding with a maximum amplitude $I_{peak}$, also shown in the bottom graph of Fig.~3. Putting everything together, with the knowledge that the input is a half-sine pulse, and expressing (2) in terms of frequency $f_m$ the equation for the input voltage of the $j^{th}$ row becomes:
\begin{equation}
V_j(t)=
\begin{cases}
	V_{0}sin\Big(2\pi f_j t+\frac{\pi}{2}\Big[1-\frac{f_j}{f_{MIN}}\Big]\Big),& \text{if } 0 \leq t \leq \frac{T_j}{2}\\
	0,	& \text{otherwise}		
	\label{eq3}
\end{cases}
\end{equation}
where $V_{0}$ is the amplitude of $V_j$, $f_{MIN}=1/T_{MAX}$ and is fixed for a full 2D convolution. As the voltage input is analogous to neuron input, $V_j(t)$ from (3) corresponds to $x_j$ from (1).

We conclude this section by noting that phase-shift is not the only mechanism by which we can align pulses. It is also possible to introduce a time-delay at the output, for example by using an \textit{RC} circuit. As we are already varying input frequency, we maintain consistency by modulating the phase instead of the output.

\section{Experimental Results}
\subsection{Weight-to-Conductance}
We experimentally demonstrate 2D convolution for image filtering in order to verify that our frequency-dependent scheme behaves correctly. We program weights in a single row of our crossbar to implement a $3\times3$ convolutional Gaussian blur smoothing filter:\\

\hbox{$\frac{1}{16}\begin{bmatrix} 1 & 2 & 1 \\ 2 & 4 & 2 \\ 1 & 2 & 1 \end{bmatrix} \implies \begin{bmatrix} 10k & 750 & 10k \\ 750 & 10 & 750 \\ 10k & 750 & 10k \end{bmatrix} \implies \begin{bmatrix} 2.1 & 4.2 & 2.1 \\ 4.2 & 8.4 & 4.2 \\ 2.1 & 4.2 & 2.1 \end{bmatrix}$} \vspace{4pt}
\noindent Here we map a kernel into input frequency in Hz, into memristor conductance in mS. As per Table I:
\begin{itemize}
\item a weight of $w=1$ corresponds to $f~=~10kHz \implies G_{ON} \approx 2.10mS$;
\item $w=2: f=750Hz \implies G_{ON}\approx 4.20mS$;
\item $w=4: f=10Hz \implies G_{ON}\approx 8.40mS$.\footnote{It is also possible to use $G_{OFF}$, and in fact will typically require lower $f$ which provides an energy saving at the cost of longer inference cycles.}
\end{itemize} 

\subsection{Image Processing}
The 2D convolution is performed using a 9$\times$9 crossbar array and a 128$\times$128 8-bit unsigned representation of a photograph of Shibuya Crossing in Tokyo from Fig.~4(a) as input with \textit{GeSeSn-W} memristors purchased from Knowm. The input image was pre-processed using artificial 8-bit RGB noise of uniform distribution with a mean of 127 and a peak of 255.


\begin{figure}[t]
\centerline{\includegraphics[scale=0.9]{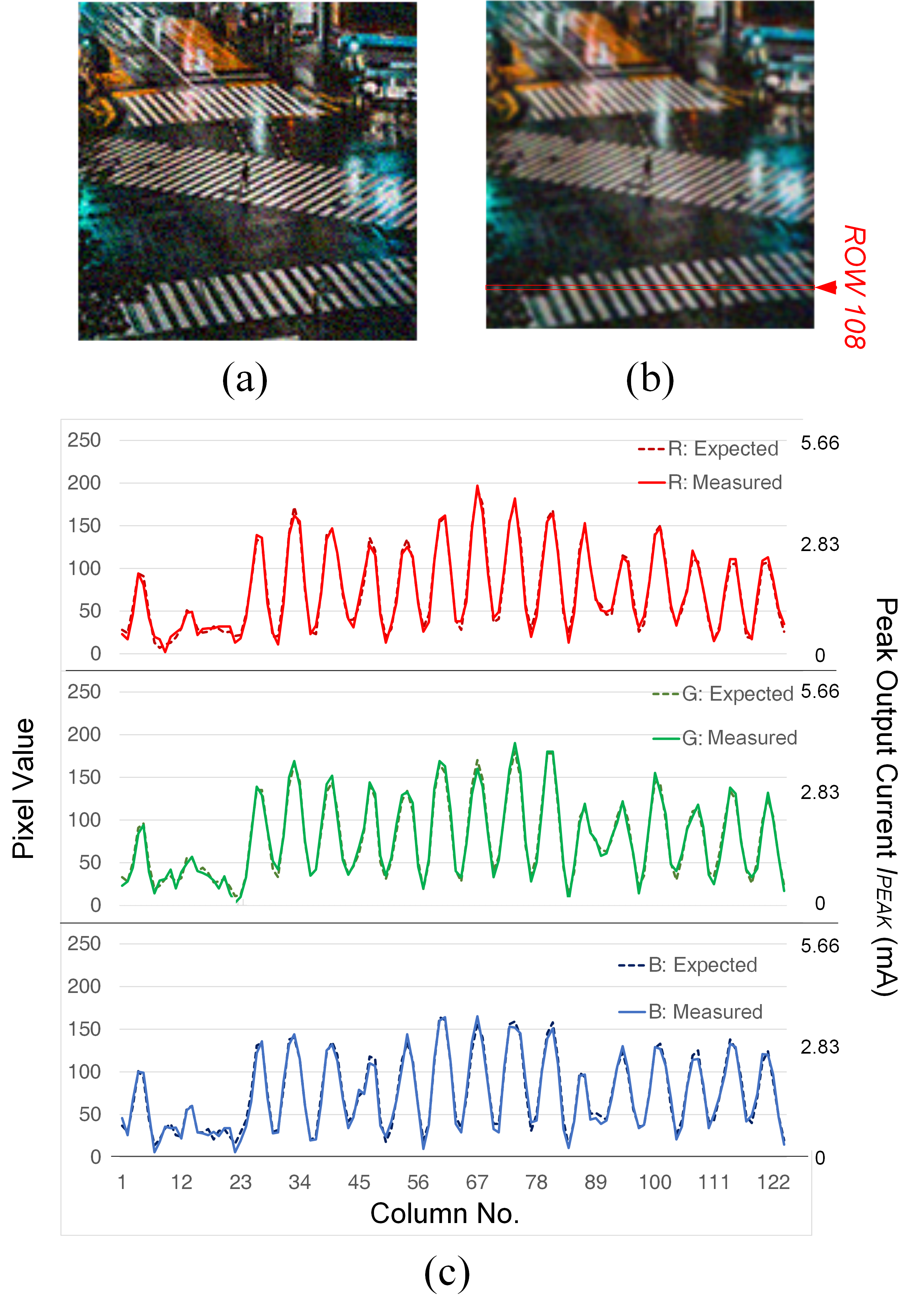}}
\caption{Gaussian blur via convolution (a) Original input image. (b) Experimental crossbar output. (c) Sample of expected and measured current peaks in terms of milliamps and corresponding bytes across row 108.} 
\label{fig4ab}
\end{figure}

With the restriction on amplitude from Section IV of $V_m(t)$ in mind, we choose $V_0 = 0.66V$ for a luminance of 255, linearly decreasing (and then modulated to frequency variation) to $V_0=0.15$ for a luminance of 0. These voltages are applied to the rows of the crossbar 9 at a time, where each output pixel is generated by the dot product of a 9-dimensional conductance vector mapped from a 3$\times$3 convolution matrix. The input image is convolved with a stride of one without zero-padding, so the dimension of the filtered image becomes 126$\times$126. We achieved this by performing 15,876 convolutions across each of the three RGB color channels.
The peak current output of each color channel is displayed in Fig.~4(c), sampled across row 108 of the photograph. The peaks and troughs in the plot correspond to the alternation between dark and light patches across the pedestrian crossing. There is good agreement between the expected values calculated by Kirchhoff's Current Law and the measured values, with the small fluctuations present due to line losses and the stochastic nature of memristors. 

Physical multiplication of a 9-dimensional vector is accomplished by a single current read process on the column wires using a Micromanipulator tungsten probe tip, with a readout time of 500 ns. This is significantly slower than the readout time in \cite{b14} which is 10 ns, though this is a limitation of the device and not the frequency-dependent method itself. In terms of the method, one limitation is that we are unable to process different filters in parallel on the same crossbar, whereas the multi-level memristor in 
\cite{b14} was shown to process 10 separate convolutional filters simultaneously. This is because the weight is now a function of not only memristor conductance, but also of input frequency. Therefore, filters cannot be distributed across different weights. This is a reasonable drawback as we are able to achieve identical processes to multi-level memristors using binarized devices. Such a limitation can always be circumvented by introducing select transistors.

\section{Discussion}
This method is shown to work successfully for integrating 8-bit operations down to a single column, thus significantly reducing area for a single convolution linearly to 12.5\% of the original size of a crossbar. This comparison does not factor the area overhead from the driving circuitry due to lack of information from prior literature, and because our input pulses and look-up table were implemented off-chip using a FPGA. On-chip implementation of input modulation is a non-trivial task, where the look-up table requires a multiplexer with select lines driven by hard-wired address signals. 

In the experimental study from the previous section, we used three weight values which were programmed using the $G_{ON}$ state. Once the operation started, there was no further need to re-write to the memristors -- we only needed a single current read process on the column wire for each convolution. Approximate average power consumption per MAC operation can be calculated using the following equation:
\begin{equation}
P=\sum_{j=0}^{m}G_j\Big(\frac{V_{j0}}{\sqrt{2}}\Big)^2,	\label{eq4}
\end{equation}
where $V_{j0}$ is the amplitude of the $j^{th}$ input. In comparison with conventional MAC crossbars such as that presented in \cite{b3}, assuming identical conductances, this method exhibits an average power saving of factor $2\times N_b$ where $N_b$ is the number of bits.\footnote{This calculation assumes ideal conditions without line resistance losses. These losses become important as the crossbar is scaled, though is negligible in our case of a 9$\times$9 crossbar and additionally explains why our expected values agree well with our measured values. The factor $2$ comes from the $(V_{RMS}/\sqrt{2})^2$ term in AC power dissipation. Sneak paths and coupling effects do not impact results as our experiments are performed with discrete components, but should be considered when scaled down to an integrated circuit \cite{b15,b16}} 
Quantitatively, for a single 2D Gaussian convolutional filter using 9 memristors along a single column, and assuming a worst case scenario where all input voltages have a maximum input peak of $V_0=0.1V$, the average power for a single MAC operation can be calculated using (4) to be 1.1mW. As a rough comparison, the process in \cite{b14} consumes $\sim$13.7mW for an image compression task although this is on a much larger 25-dimension voltage vector (i.e., 5$\times$5 kernel). Regardless, it is evident this method is competitive with current state-of-the-art processes, whilst using devices that operate on much larger conductances. Therefore, there is good potential for further power savings if crossbars that operate with micro-scale conductances implement this frequency-dependent strategy.

\section{Conclusion}
This paper has presented a frequency-based mechanism for using binarized-conductance switching to generate analog weights, which is shown to reduce power when compared to column-distribution methods (as partially characterized by (4)), and to also reduce area by a factor equivalent to the number of bits used in image representation. This is performed without the need for multi-level memristors so can be implemented using phase-change memory switching or metal-oxide resistive switches, and is immaterial to the physical composition of the devices.\\ 
Importantly,  we use a generalizable method that can implement different look-up tables by using different devices. While binarized NNs are known to perform well for most simple classification tasks, there is ultimately some compromise with accuracy that we are able to avoid using our anolog approach \cite{b17}.\\
We achieve a highly efficient method for signal and image processing using convolution, and improvements in MAC operations in DNNs and CNNs. Future work includes testing this scheme on different devices to calculate read stability across various devices, and developing a way to pipeline frequency-dependency across columns to decrease latency.


\section*{Acknowledgements}
\noindent This work was supported by the Australian Department of Foreign Affairs and Trade, Australia-Korea Foundation under Grant AKF00640, the Commonwealth Government of Australia through the Australian Government Research Training Program Scholarship, and iDataMap Corporation. 

\end{document}